\documentclass[10pt]{article}
\input xy
\xyoption{all}
\usepackage{amsthm}
\usepackage{amsfonts,amsmath}

\def\tp{\otimes} 
\def\mod{\hspace{0.2cm} \mbox{mod} \hspace{0.2cm}} 
\def\modn{\hspace{0.2cm} (\mbox{mod} \hspace{0.2cm} n)} 

\def\d{\delta}

\def\N{\mathbb{N}}
\def\Z{\mathbb{Z}}

\def\R{\mathbb{R}}
\def\C{\mathbb{C}}
\def\L{\mathcal{L}}

\theoremstyle{definition}

\theoremstyle{remark}

\title{Integrable Hamiltonians with $D(D_n)$ symmetry from the Fateev-Zamolodchikov model}
\author{P.E. Finch \\
  \\
  Institut f\"ur Theoretische Physik, \\
  Leibniz Universit\"at Hannover, \\
  Appelstra\ss e 2, 30167 Hannover, Germany}
\date{}

\begin{document}
\maketitle

\begin{abstract}
\noindent
A special case of the Fateev-Zamolodchikov model is studied resulting in a solution of the Yang-Baxter equation with two spectral parameters. Integrable models from this solution are shown to have the symmetry of the Drinfeld double of a dihedral group. Viewing this solution as a descendant of the zero-field six-vertex model allows for the construction of functional relations and Bethe ansatz equations.
\end{abstract}

\section{Introduction}

\noindent
The connetion between the Yang--Baxter equation and integrable systems has been well-studied with origins in the works of McGuire, Yang and Baxter \cite{BaxterBook1982,McGuire1966,Yang1967}. 
Solutions of the Yang--Baxter equation, $R$-matrices, allow the construction of integrable one-dimensional quantum chains via the quantum inverse scattering method (QISM) \cite{STF1979}. Typically the quantum chains produced consist of nearest neighbour interactions, are translationally invariant and have periodic boundary conditions.
However, it is possible to modify this procedure resulting in a model with \textit{twisted} boundary conditions \cite{Vega1984,FLR1998,ShaSut1990}, in which periodicity is retained but translational invariance is broken as the interaction between the \textit{first} and \textit{last} site is distinct from interactions between other neighbours.
The periodicity of the model can also be broken using Skylanin's approach \cite{Sklyanin1988}; this procedure was termed BQISM (boundary quantum inverse scatter method) and leads to quantum chains with \textit{open} boundary conditions. Skylanin's approach required the advent of \textit{reflection matrices} and assumed multiple properties of the $R$-matrix involved. Over time these assumed properties have been systematically removed \cite{BGZZ1998,LinGou1996,MezNep1991,MezNep1992,Zhou1996}. 
Operators which lead to a quantum chain with open boundaries can be modified to produce a quantum chain with \textit{braided closed} boundary conditions \cite{Foerster1996,GPPR1994,KarZap1994,LinFoe1997,LFK1999,LimaSantos1998}.

Connected to the Yang--Baxter equation are the algebraic structures known as quasi-triangular Hopf algebras. These structures are known to provide algebraic solutions of the Yang--Baxter equation, albeit without so-called spectral parameters. The addition of spectral parameters can be achieved through Baxterisation \cite{CGX1991,Jones1990}. The Drinfeld doubles of finite group algebras are one notable class of quasi-triangular Hopf algebras \cite{DPR1990}. 
Although attention has recently been paid to these algebras in relation to non-abelian anyonic theories, not much interest has been directed towards the connection between their associated $R$-matrices and integrable quantum chains.
We remark that in contrast to the more familiar case that the $R$-matrix comes from a quantum group, the quantum chain is interpreted as an anyon chain \cite{CDIL2010,FTLTKWF2007} rather than a spin chain.

We are concerned with the Fateev--Zamolodchikov model \cite{FatZam1982b}, which is known to be a limiting case of the chiral Potts model \cite{AYMCPTY1987}. The Fateev--Zamolodchikov model has received attention for both open boundary conditions \cite{Zhou1997} and periodic boundary conditions \cite{Albertini1992,AMCPT1989,RaySha2005}. The latter works are restricted to the \textit{uniform square lattice} case of the checkerboard lattice model \cite{BPAY1988,PerAuY2006}. This special limiting case is found by making certain rapidities equal. We however are concerned with an alternate limit of the Fateev--Zamolodchikov model. In this limit we perform a basis transformation on the $R$-matrix, again constructed as checkerboard vertex model, where the symmetry of $D(D_{n})$ is recognisable. We discuss three different classes of boundary conditions for quantum chains, namely, periodic, open and braided closed boundary conditions. Lastly we present the $R$-matrix as a descendant of the six-vertex model and construct some fusion relations for the models discussed.

\section{Preliminaries}
Throughout $\delta_{i}^{j}$ will represent the Kronecker delta function, being one if $i=j$ and zero otherwise. We do not restrict $i$ and $j$ to the integers and they may belong to any group.

We define $e_{i,j}$ to be an elementary matrix of $M_{d \times d}(\C)$ whose indices are considered modulo $d$. These matrices obey the relation
$$ e_{i,j}e_{k,l} = 
	\begin{cases}
		e_{i,l}, & ~~ j \equiv k \mod d, \\
		0, & ~~ j \not\equiv k \mod d.
	\end{cases}
$$
Our convention is that $e_{i,j}$ corresponds to the matrix with a one in the $i$th row and $j$th column and zeros elsewhere for $1 \leq i,j \leq d$. 
For such operators we define the set of $d$-dimensional vectors $\{v_{k}\}_{k}$ satisfying the property
$$ e_{j,k}v_{k} = v_{j}, \hspace{1cm} \forall j,k \in \Z. $$
Under this definition we have that the indices of the vectors are considered modulo $d$ as is the case with the elementary matrices. \\

\noindent
\underline{\textbf{The Yang--Baxter equation}}\\
\noindent
The Yang--Baxter equation (YBE) is a non-linear equation in the three fold tensor product of some algebra $A$. There are multiple variants of this equation. The first we consider is the \textit{constant} YBE, given by
$$ R_{12} R_{13} R_{23} = R_{23} R_{13} R_{12}, $$
where the indices of the operator $R \in A \tp A$ indicate the spaces which the operator acts non-trivially on. That is,
$$ R_{12} = R \tp I, \hspace{0.5cm} R_{23} = I \tp R, \hspace{0.5cm} \mbox{etc.,} $$
where $I$ is the identity element of $A$. The \textit{parameter} dependant form of the YBE is given by
\begin{equation}  \label{eqnYBE}
R_{12}(\tilde{x};\tilde{y}) R_{13}(\tilde{x};\tilde{z})R_{23}(\tilde{y};\tilde{z}) = R_{23}(\tilde{y};\tilde{z}) R_{13}(\tilde{x};\tilde{z}) R_{12}(\tilde{x};\tilde{y}),
\end{equation}
where $\tilde{x}$, $\tilde{y}$ and $\tilde{z}$ are elements of the Cartesian product of $d$ copies of $\C$, which we consider a group under multiplication. This equation can be simplified if the $R$-matrix satisfies the multiplicative analogue of the so-called \textit{difference property}, 
\begin{equation} \label{eqnYBEdiffpara}
	R_{12}(\tilde{x}) R_{13}(\tilde{x}\tilde{y})R_{23}(\tilde{y}) = R_{23}(\tilde{y}) R_{13}(\tilde{x}\tilde{y}) R_{12}(\tilde{x}).
\end{equation}
Solutions to any of these equation will be referred to as $R$-\textit{matrices}.

When the $R$-matrix is a matrix operator, i.e $R(\tilde{z}) \in \mbox{End}(V \tp V)$ for some vector space $V$, we use the definitions of regularity and unitarity,
$$ R(\tilde{1}) = P \hspace{0.7cm} \mbox{and} \hspace{0.7cm} R_{12}(\tilde{z})R_{21}(\tilde{z}^{-1}) \propto I \tp I, $$
where $P$ is the usual permutation operator
$$ P(v \tp w) = w \tp v, \hspace{1cm} v,w \in V. $$
We assume every matrix operator $R$-matrix satisfies regularity, unless otherwise stated. It is known that regularity implies unitarity. \\

\noindent
\underline{\textbf{The Drinfeld doubles of dihedral groups}}\\
\noindent
The algebraic structures which we focus upon are the Drinfeld doubles of dihedral groups. The general structure, along with the representation theory, of the Drinfeld doubles of finite group algebras is known \cite{DPR1990,Gould1993}. For any finite group, $G$, the defining relations of its Drinfeld double, $D(G)$, can be presented purely in terms of the operation of the group. There are procedures which allow the construction of all irreducible representations (irreps) of $D(G)$.

We are interested in the dihedral groups of order $2n$, where $n$ is odd. The dihedral group $D_{n}$ corresponds to the symmetries of a regular polygon with $n$ vertices. It has the presentation 
$$ D_{n} = \{\sigma, \tau | \sigma^{n} = \tau^{2} = \sigma \tau \sigma \tau = e \}, $$
where $e$ is the identity element of the group. We now consider the Drinfeld double of this group. It is defined as the vector space:
$$ D(D_{n}) = \C \{gh^{*} | g,h \in D_{n}\}, $$
equipped with the multiplication and coproduct
$$ g_{1}h_{1}^{*} g_{2}h_{2}^{*} =  \d_{(h_{1}g_{2})}^{(g_{2}h_{2})} \,\, (g_{1}g_{2}) h_{2}^{*} \hspace{0.5cm} \mbox{and} \hspace{0.5cm} \Delta(gh^{*}) = \sum_{k\in D_{n}} g (k^{-1}h)^{*} \tp g k^{*}. $$
The remaining Hopf structure is uniquely determined by these relations. This algebra is associated with a canonical element,
$$ \mathfrak{R} = \sum_{g\in D_{n}} g \tp g^{*}, $$
which is known to satisfy the YBE. 
We define the twist map and twisted coproduct to be,
$$ T(a\tp b) = b \tp a \hspace{0.5cm} \mbox{and} \hspace{0.5cm} \Delta^{T}(a) = T \circ \Delta(a), \hspace{0.5cm} \forall a,b\in D(D_{n}), $$
respectively. The $R$-matrix satisfies
$$ \mathfrak{R} \Delta(a) = \Delta^{T}(a)\mathfrak{R} $$
for all $a \in D(D_{n})$.

The explicit irreps of $D(D_{n})$ have already been constructed \cite{DIL2006}, however, we shall only use one of dimension $n$. Firstly we view $D_{n}$ as a subgroup of $S_{n}$, the group of permutations on $n$ objects. This affords $D_{n}$ an action on $\Z_{n}$, the ring of integers modulo $n$. This action is defined by
$$ \sigma = [1,2,...,n] \hspace{0.7cm} \mbox{and} \hspace{0.7cm} \tau = \prod_{k=1}^{\frac{n-1}{2}}[k,n-k], $$
using the standard notation of $S_{n}$ \cite{FraleighBook1982}. Thus $\sigma$ increments an integer by one and $\tau$ maps an integer to its negative, both modulo $n$. This action of $D_{n}$ allows us to present the $n$-dimensional irrep in the following manner:
$$ \pi(g) = \sum_{i=1}^{n} e_{g(i),i} \hspace{0.7cm} \mbox{and} \hspace{0.7cm} \pi(g^{*}) = \sum_{j=0}^{n-1} \delta_{g}^{\sigma^{2j}\tau}e_{j,j}. $$
This representation acts on the space $\C^{n}$ and can be naturally extended to $(\C^{n})^{\tp l}$ through the definition
$$ \pi_{n}^{\otimes l}(a) \equiv \pi_{n}^{\otimes l} \left(\Delta^{(l)}(a)\right), \hspace{1cm} \forall a \in D(D_{n}) $$
where $\Delta^{(l)}$ is defined recursively by the relations
$$ \Delta^{(k)} = (\Delta \tp \mbox{id}^{\otimes (k-2)})\Delta^{(k-1)} \hspace{0.7cm} \mbox{and} \hspace{0.7cm} \Delta^{(2)} = \Delta, $$
for $k \geq 3$, with $\mbox{id}$ being the identity map. \\

\noindent
\underline{\textbf{The Fateev--Zamolodchikov model}}\\
\noindent
The models we consider are special cases of the well-known Fateev--Zamolodchikov model \cite{FatZam1982b}. The Fateev--Zamolodchikov model is defined by weights
\begin{eqnarray}
	W(z|0) = 1, && W(z|l) = \prod_{j=1}^{l} \frac{\lambda^{2j-1}z-1}{\lambda^{2j-1}-z},  \nonumber\\
	\overline{W}(z|0) = 1\phantom{,} & \hspace{0.7cm}\mbox{and}\hspace{0.7cm} & \overline{W}(z|l) = \prod_{j=1}^{l} \frac{\lambda^{2j-1} - \lambda z}{\lambda^{2j}z  - 1}, \nonumber 
\end{eqnarray}
where $\lambda$ is a primitive $2n$th root of unity and $z \in \C$ and $0 \leq l \leq n-1$. These weights satisfy the star-triangle relation\footnote{See \cite{PerAuY2006} for definition of the star-triangle relation.} and are extended by the relations
$$ W(z|l) = W(z|n+l) = W(z|-l) \hspace{0.7cm} \mbox{and} \hspace{0.7cm} \overline{W}(z|l) = \overline{W}(z|n+l) = \overline{W}(z|-l), $$
for $l\in \N$. Using these weights we have the Fateev--Zamolodchikov $R$-matrix
$$ R(\tilde{x};\tilde{y}) = \sum_{a,b,c,d=1}^{n} \overline{W}(x_{1}y_{1}^{-1}|b-c) W(x_{2}y_{1}^{-1}|b-d) \overline{W}(x_{2}y_{2}^{-1}|a-d) W(x_{1}y_{2}^{-1}|a-c) e_{a,b} \tp e_{c,d}. $$
The weights here have been configured in a checkerboard lattice configuration. The uniform square lattice arises when the two different rapidities are set equal, $x_{1}=x_{2}$ and $y_{1}=y_{2}$ \cite{BPAY1988}.

\section{Integrable $n$-state models}
An observation of Bazhanov and Perk was that the $D(D_{3})$ model presented in \cite{DIL2006} was a special case of the Fateev--Zamolodchikov model \cite{BazPer2009}. Moreover it was implied that all the $R$-matrices derived from $D(D_{n})$ as presented in \cite{FDIL2011} should also be special cases as was shown to be the case for odd $n$. Here we show that there is in fact a more general limit of the Fateev--Zamolodchikov $R$-matrix that has the underlying symmetry of $D(D_{n})$.

We take a limit of the Fateev--Zamolodchikov $R$-matrix and define
$$ R(\tilde{z}) = \lim_{x_{2},y_{2}\rightarrow 0} \lim_{x_{1},y_{1}\rightarrow \infty} R(\tilde{x};\tilde{y}), $$
where $z_{1} = \frac{x_{1}}{y_{1}}$ and $z_{2} = \frac{y_{2}}{x_{2}}$. Explicitly we have
$$ R(\tilde{z}) = \sum_{a,b,c,d=1}^{n} (-1)^{a+b+c+d} \lambda^{(a-c)^{2}-(b-d)^{2}} \overline{W}(z_{1}|b-c) \overline{W}(z_{2}^{-1}|a-d) e_{a,b} \tp e_{c,d}. $$
This has the properties
$$ R(1,1) = P \hspace{0.7cm} \mbox{and} \hspace{0.7cm} PR(z_{1},z_{2})P = R(z_{2}^{-1},z_{1}^{-1}), $$
This is now an $R$-matrix which satisfies the variant of the YBE given by Equation (\ref{eqnYBEdiffpara}). To put the $R$-matrix in a recognisable form we perform the basis transformation corresponding to the matrices
$$ S = \frac{1}{\sqrt{n}} \sum_{i,j=1}^{n}\lambda^{2j(1-j)} e_{2(i+j),i} \hspace{0.7cm} \mbox{and} \hspace{0.7cm} S^{-1} = \frac{1}{\sqrt{n}} \sum_{i,j=1}^{n} \lambda^{2j(j-1)}e_{i,2(i+j)}. $$
The $R$-matrix becomes
\begin{eqnarray}
	R(\tilde{z}) 
	& = & \sum_{a,i,j=1}^{n} \left[\sum_{b=1}^{n} w^{-2a(2b-j)} \overline{W}(z_{1}|b) \overline{W}(z_{2}^{-1}|b-j) \right] e_{i+j,i+a} \tp e_{i+a+j,i}, \label{DDnRmat}
\end{eqnarray}
where we have replaced $\lambda$ with $-w^{-2}$, with $w$ being a primitive $n$ root of unity. We can express the $R$-matrix in terms of the projection operators presented in \cite{FDIL2011}. The projection operators are
\begin{eqnarray*}
	p^{(a,b)} & = & \frac{c^\alpha}{n} \sum_{i,j = 0}^{n-1}[ w^{2bj} e_{i+a+j,i+a} \tp e_{i+j,i} + w^{-2bj} e_{i-a+j,i-a} \tp e_{i+j,i}], \hspace{0.9cm} \mbox{where},  \\
 	c^{(a,b)} & = &\begin{cases} \frac{1}{2}, &\quad (a,b) = (0,0), \\ 1, & \quad (a,b) \neq (0,0). \end{cases}
\end{eqnarray*}
The admissible pairs, $(a,b)$, for the projection operators along with the corresponding irrep they project onto can be found in Table 2 of \cite{FDIL2011}. Using these projection operators we can express the $R$-matrix as
\begin{eqnarray*}
	P R(\tilde{z})
	& = & \sum_{b=0}^{\frac{n-1}{2}} f_{(0,b)}(\tilde{z}) p^{(0,b)} + \sum_{a=1}^{\frac{n-1}{2}} \sum_{b=0}^{n-1} f_{(a,b)}(\tilde{z}) p^{(a,b)},
\end{eqnarray*}
where 
\begin{eqnarray}
	f_{(a,b)}(\tilde{z})
	& = & \left[\sum_{c=1}^{n} w^{2(a+b)c} \overline{W}(z_{1}|c) \right] \left[ \sum_{d=1}^{n}w^{2(a-b)d}\overline{W}(z_{2}^{-1}|d) \right] \label{eqnfab}.
\end{eqnarray}
As the $R$-matrix can be written as linear combination of the above projection operators in such a manner it follows that the $R$-matrix intertwines the coproduct of $D(D_{n})$. We express this as
$$ \left[ P R(\tilde{z}), \, \pi^{\tp 2}(a) \right] = 0, \hspace{1cm} \forall a \in D(D_{n}). $$
It is known that setting $z_{1}=z_{2}=z$ recovers the $R$-matrix presented in \cite{FDIL2011}. Looking at the functions $f_{(a,b)}(\tilde{z})$ we see that they factorise into functions of $z_{1}$ and $z_{2}$ and that the components involving $z_{2}$ are determined purely by the difference $a-b$. Both of these are required by the construction of the $D(D_{n})$ $R$-matrix as a descendant presented in \cite{FDIL2011}.

In this basis we also have the additional property
$$ \left[R(z_{1},z_{2})\right]^{*} = R(z_{2}^{*},z_{1}^{*}). $$
By investigation of the non-zero entries of the $R$-matrix we see that there is a generalised eight-vertex condition present. However, this generalisation differs from others \cite{AuYPer1997} which stems from different conservation rules. The conservation rule which is satisfied here is 
$$ R_{ij}^{kl}(\tilde{z}) = 0 \hspace{0.3cm} \mbox{when} \hspace{0.3cm} i+j \neq k+l \modn \hspace{0.5cm} \mbox{for the operator} \hspace{0.5cm} R(\tilde{z}) = \sum_{i,j,k,l=1}^{n}  R_{ij}^{kl}(\tilde{z}) e_{i,j} \tp  e_{k,l}. $$

Important to constructing integrable quantum chains are the logarithmic derivatives\footnote{We define the logarithmic derivative as: $\frac{d}{dz}\ln[A(z)] = A^{-1}(z)\frac{d}{dz}[A(z)]$} of $R(\tilde{z})$. We define
$$ H^{(1)} = i\left[\frac{d}{dz_{1}} \ln(R(\tilde{z}))\right]_{\tilde{z}=\tilde{1}} \hspace{0.5cm} \mbox{and} \hspace{0.5cm} H^{(2)} = -i\left[\frac{d}{dz_{2}} \ln(R(\tilde{z}))\right]_{\tilde{z}=\tilde{1}}. $$ 
Using these definitions we have
\begin{eqnarray}
	H^{(1)}
	& = & i \sum_{i,j=1}^{n} \sum_{l=1}^{n-1} \left[ (-1)^{l} \frac{w^{-2l(i-j)}}{\left(w^{2l} - w^{-2l}\right)}\right] e_{i+l,i} \tp e_{j+l,j} \nonumber \\
	& = & i \sum_{a=1}^{\frac{n-1}{2}} \sum_{b=0}^{n-1} \left[ (-1)^{a} \frac{w^{2ab}}{w^{2a} - w^{-2a}} \right] \sum_{\gamma \in D_{n}}e_{\gamma(a-b),\gamma(n-b)} \tp e_{\gamma(a),\gamma(n)} \nonumber 
\end{eqnarray}
and, using $H^{(2)} = PH^{(1)}P$,
\begin{eqnarray}
	H^{(2)}
	& = & i \sum_{i,j=1}^{n} \sum_{l=1}^{n-1} \left[ (-1)^{l} \frac{w^{2l(i-j)}}{\left(w^{2l} - w^{-2l}\right)}\right] e_{i+l,i} \tp e_{j+l,j} \nonumber \\
	& = & i \sum_{a=1}^{\frac{n-1}{2}} \sum_{b=0}^{n-1} (-1)^{a} \left[ \frac{w^{-2ab}}{w^{2a} - w^{-2a}} \right] \sum_{\gamma \in D_{n}} e_{\gamma(a-b),\gamma(n-b)} \tp e_{\gamma(a),\gamma(n)} \nonumber 
\end{eqnarray}
These two operators are self-adjoint and complex conjugates
$$ \left[H^{(1)}\right]^{\dagger} = H^{(1)} \hspace{0.7cm} \mbox{and} \hspace{0.7cm} \left[ H^{(2)} \right]^{*} = H^{(1)}. $$
We have that the Eigenvalues of $H^{(1)}$ and $H^{(2)}$ are real, furthermore, the Eigenspectrums of $H^{(1)}$ and $H^{(2)}$ must be the same.

These operators will form the local Hamiltonians of the quantum chains discussed in the next section. 
Although it is possible to present them in terms of spin operators, there is no physical motivation for this as the models will lack the usual underlying spin symmetries. Instead we see the emergence of the $D(D_{n})$ symmetry in these operators with summation over the elements of $D_{n}$, implying the known invariance. The coefficients of the fusions channels, indicating energetically favoured paths, can be calculated from the functions given in Equation (\ref{eqnfab}) \cite{NSSFDS2008,TTWL2009}. \\

\subsection{Quantum chains}
Here we will discuss different models which can be constructed. To describe the models we consider a general framework, in which we have combined quantum chains with periodic, twisted, open and braided closed boundary conditions. For this we replace the single $R$-matrix with four operators, noting that for each class of model considered the different operators will be trivially related to the original $R$-matrix or themselves trivial. The four operators we consider are those appearing in the equations, which they must satisfy, below,
\begin{eqnarray}
	R_{12}(\tilde{x}) R_{13}(\tilde{x}\tilde{y}) R_{23}(\tilde{y}) & = & R_{23}(\tilde{y}) R_{13}(\tilde{x}\tilde{y}) R_{12}(\tilde{x}), \label{YBE1} \\
	R_{12}'(\tilde{x}) \bar{R}_{13}(\tilde{x}\tilde{y}) \bar{R}_{23}(\tilde{y}) & = & \bar{R}_{23}(\tilde{y}) \bar{R}_{13}(\tilde{x}\tilde{y}) R_{12}'(\tilde{x}), \label{YBE2} \\
	\bar{R}_{13}(\tilde{x}) \bar{R}_{12}(\tilde{x}\tilde{y}) R_{23}(\tilde{y}) & = & R_{23}(\tilde{y}) \bar{R}_{12}(\tilde{x}\tilde{y}) \bar{R}_{13}(\tilde{x}), \label{YBE3} \\
	R_{13}(\tilde{x}) \bar{R}_{12}'(\tilde{x}\tilde{y}) \bar{R}_{23}(\tilde{y}) & = & \bar{R}_{23}(\tilde{y}) \bar{R}_{12}'(\tilde{x}\tilde{y}) R_{13}(\tilde{x}). \label{YBE4}
\end{eqnarray}
The first of these equations is the usual aforementioned Yang--Baxter equation while the last three are simple variants. We also introduce two reflection equations along with reflection matrices
\begin{eqnarray}
	R_{12}(\tilde{x}\tilde{y}^{-1}) K_{1}^{-}(\tilde{x}) \bar{R}_{12}(\tilde{x}\tilde{y}) K_{2}^{-}(\tilde{y}) & = & K_{2}^{-}(\tilde{y}) \bar{R}_{12}'(\tilde{x}\tilde{y}) K_{1}^{-}(\tilde{x}) R_{12}'(\tilde{x}\tilde{y}^{-1}), \label{RE1} \\
	\left[R_{12}'(\tilde{x}\tilde{y}^{-1})\right]^{-1} K_{1}^{+}(\tilde{x})\tilde{R}_{12}'(\tilde{x}\tilde{y}) K_{2}^{+}(\tilde{y}) & = & K_{2}^{+}(\tilde{y}) \tilde{R}_{12}(\tilde{x}\tilde{y}) K_{1}^{+}(\tilde{x}) \left[R_{12}(\tilde{x}\tilde{y}^{-1})\right]^{-1},\label{RE2}
\end{eqnarray}
where
$$ \left[\tilde{R}_{12}(\tilde{z})\right]^{t_{1}} \left[ \bar{R}_{12}(\tilde{z}) \right]^{t_{1}} =  I \tp I \hspace{0.7cm} \mbox{and} \hspace{0.7cm} \left[\tilde{R}_{12}'(\tilde{z})\right]^{t_{2}} \left[\bar{R}_{12}'(\tilde{z})\right]^{t_{2}} = I \tp I. $$
We have assumed the invertibility of $\bar{R}_{12}^{t_{1}}(\tilde{z})$ and $\bar{R}_{12}^{t_{2}}(\tilde{z})$ which is not always guaranteed. Provided that the above six equations are satisfied then it follows that the monodromy matrix
$$ T_{0(1..\L)}(\tilde{z}) = R_{0\L}(\tilde{z})...R_{01}(\tilde{z})K_{0}^{-}(\tilde{z})\bar{R}_{01}(\tilde{z})...\bar{R}_{0\L}(\tilde{z}), $$
satisfies
\begin{equation}
R_{12}(\tilde{x}\tilde{y}^{-1}) T_{13}(\tilde{x}) \bar{R}_{12}(\tilde{x}\tilde{y}) T_{23}(\tilde{y}) = T_{23}(\tilde{y}) \bar{R}_{12}'(\tilde{x}\tilde{y}) T_{13}(\tilde{x}) R_{12}'(\tilde{x}\tilde{y}^{-1}). \nonumber 
\end{equation}
This leads the transfer matrix
$$ t(\tilde{z}) = \mbox{tr}_{0} \left[ K^{+}_{0}(\tilde{z}) T_{0(1..\L)}(\tilde{z}) \right], $$
which is guaranteed to satisfy
$$ [t(\tilde{x}), t(\tilde{y})] = 0. $$
This commutation relation ensures the integrability of the Hamiltonian:
$$ \mathcal{H} = \sum_{j=1}^{d} \alpha_{j}\beta_{j} \left\{ \left[\frac{d}{dz_{j}} \ln(t(\tilde{z}))\right] - \left[\mbox{dim}(V)\right]^{-\L} \mbox{tr}\left[\frac{d}{dz_{j}} \ln(t(\tilde{z}))\right] I^{\tp \L} \right\}_{\tilde{z}=1}, $$
where $\alpha_{j} \in \R$ are free and $\beta_{j} \in \C$ are fixed. The later scalars are chosen and fixed to ensure self-adjointness when possible. We have also included a term to make the trace of the global Hamiltonian zero.

For each model we will be able to write the global Hamiltonians in terms of other Hamiltonians acting on the entire space. That is,
$$ \mathcal{H} = \sum_{j=1}^{d} \alpha_{j} \mathcal{H}^{(j)}, $$
with each of the $\mathcal{H}^{(j)}$ and $\mathcal{H}^{(k)}$ commuting. Instead of calculating the Eigenspectrum of $\mathcal{H}$, often the individual spectra of the $\mathcal{H}^{(j)}$ will be calculated. However, to recover the spectrum of the complete Hamiltonian so-called \textit{pairing} or \textit{grouping} rules will need to be found. 
Generally the ground state (and its energy) will depend upon the $\alpha_{j}$ with level crossings and quantum phase transitions expected. \\

\noindent
\underline{\textbf{Periodic and twisted boundary models}}\\
\noindent
To construct a periodic model we set
$$ R'(\tilde{z}) = R(\tilde{z}) \hspace{0.5cm} \mbox{and} \hspace{0.5cm} \bar{R}(\tilde{z})=\bar{R}'(\tilde{z}) = I \tp I. $$
We immediately find Equations (\ref{YBE2}-\ref{YBE4}) are trivially satisfied. Furthermore without loss of generality we can set $K^{+}(\tilde{z})=I$ satisfying Equation (\ref{RE2}). The only operator left to discuss is $K^{-}(z)$ which must satisfy
$$ R(\tilde{x}\tilde{y}^{-1}) (K^{-}(\tilde{x}) \tp K^{-}(\tilde{y})) = (K^{-}(\tilde{x}) \tp K^{-}(\tilde{y})) R(\tilde{x}\tilde{y}^{-1}). $$
This equation can clearly always be satisfied although the spectrum of solutions depends heavily on $R(\tilde{z})$. The transfer matrix simplifies down to
$$ t(\tilde{z}) = \mbox{tr}_{0} \left[ K_{0}^{-}(\tilde{z}) R_{0\L}(\tilde{z})...R_{01}(\tilde{z}) \right]. $$
In the case that $K^{-}(\tilde{z})$ is independent of $\tilde{z}$ the model presented is precisely the definition of a periodic model with twisted boundaries as defined in \cite{FLR1998}. For the $R$-matrices we are concerned with there is a class of such solutions. For each group element $g \in D_{n}$ we can set
$$ K^{-}(z) = \pi(g). $$
The global Hamiltonian will be
\begin{equation} \label{eqnGloHtwist}
	\mathcal{H} = \sum_{i=1}^{\L -1} H_{i(i+1)} +  H^{g}_{\L 1} 
\end{equation}
where
$$ H = \sum_{j=1}^{d} \alpha_{j}\beta_{j} \left\{ \left[\frac{d}{dz_{j}} \ln(R(z))\right] - \mbox{tr}\left[\mbox{dim}(V)\right]^{-2} \mbox{tr}\left[\frac{d}{dz_{j}} \ln(R(z))\right] I \tp I \right\}_{\tilde{z}=1}, $$
is the local Hamiltonian and
$$ H^{g} = (\pi(g)^{-1} \tp I) H (\pi(g)  \tp I), $$
is the twisted local Hamiltonian. Explicitly we have for the $R$-matrix given by Equation (\ref{DDnRmat}) that the local Hamiltonian is a linear combination of the operators $H^{(1)}$ and $H^{(2)}$,
\begin{equation}
	H = i \sum_{a=1}^{\frac{n-1}{2}} \sum_{b=0}^{n-1} \left[ (-1)^{a} \frac{\alpha_{1}w^{2ab} + \alpha_{2}w^{-2ab}}{w^{2a} - w^{-2a}} \right] \sum_{\gamma \in D_{n}}e_{\gamma(a-b),\gamma(n-b)} \tp e_{\gamma(a),\gamma(n)}, \label{Hlocal}
\end{equation}
while the twisted local Hamiltonian is
\begin{equation}
	H^{g} = i \sum_{a=1}^{\frac{n-1}{2}} \sum_{b=0}^{n-1} \left[ (-1)^{a} \frac{\alpha_{1}w^{2ab} + \alpha_{2}w^{-2ab}}{w^{2a} - w^{-2a}} \right] \sum_{\gamma \in D_{n}}e_{\gamma(a-b),\gamma(n-b)} \tp e_{g \circ \gamma(a),g \circ \gamma(n)}. \nonumber
\end{equation}
The global Hamiltonian is self-adjoint.

Setting $g=e$ recovers a periodic chain, while additionally setting $\alpha_{1}=1=-\alpha_{2}$ yields the Hamiltonian for the $R$-matrix described in \cite{FDIL2011}. We have the property that the local untwisted Hamiltonian commutes with both the coproduct and twisted coproduct, that is,
$$ [H_{12}, (\pi \tp \pi)\Delta(a)] = [H_{21}, (\pi \tp \pi)\Delta(a)] = 0, \hspace{1cm} \forall a \in D(D_{n}). $$
However, we find that for periodic chains of finite length the global Hamiltonian does not inherit the full symmetry of $D(D_{n})$. In this case we find that the inherited symmetry of the global Hamiltonian is instead given by
$$ C_{\mathcal{H}} = \left\{a\in D(D_{n})| \, \Delta(a)=\Delta^{T}(a) \right\}. $$
That is,
$$ [\mathcal{H}, (\pi_{n}^{\otimes \L})(a)] = 0, \hspace{1cm} a \in C_{\mathcal{H}}. $$
The periodic chain of any length will be translationally invariant as
$$ t(\tilde{1}) = P_{1\L}...P_{13}P_{12}, $$
must commute with the Hamiltonian.

For general $g$ we can see that the inherited symmetry of the system is reduced further. 
The global Hamiltonian will not be translationally invariant. However, we can consider
$$ t(\tilde{1}) = K_{1}^{-}(1) P_{12}P_{23}..P_{(\L-1)\L} $$
to play the role of a generalised translation operator. 
To now leave the model unchanged we need to perform a translation, i.e. map site $i$ to site $i+1$ modulo $n$, and additionally perform a transformation on the $1$st site.
We also remark that if we have group elements $g$ and $h$ which are in the same conjugacy class then the global Hamiltonians produced from both are equivalent. \\

\noindent
\underline{\textbf{Open boundary models}}\\
\noindent
To construct an open model we set
$$ R_{12}'(\tilde{z}) = \bar{R}_{12}(\tilde{z}) = R_{21}(\tilde{z}) \hspace{0.5cm} \mbox{and} \hspace{0.5cm} \bar{R}_{12}'(\tilde{z})= R_{12}(\tilde{z}). $$
We find that with these definitions Equations (\ref{YBE1}-\ref{YBE4}) are all equivalent and both (\ref{RE1},\ref{RE2}) are non-trivial. One particular case is when the $K$-matrices are the identity i.e.
$$ K^{-}(\tilde{z}) = K^{+}(\tilde{z}) = I. $$
This is always possible for the $R$-matrix presented. This gives the global Hamiltonian
$$ \mathcal{H} = \sum_{i=1}^{\L -1} H_{i(i+1)}, $$
where $H$ proportional to the local Hamiltonian presented in the periodic case. This system is described as having non-interacting boundary terms or free ends due to the lack of operators acting solely on sites $1$ or $\L$. Unlike the periodic case the global Hamiltonian inherits the complete symmetry of $D(D_{n})$ for finite size chains. The introduction of interacting boundary terms will break this symmetry. 
We find the model has no form of translational invariance and that the operator $t(\tilde{1})$, which we previously identified with translation, is a scalar multiple of the identity. The latter property of the model also implies that the logarithmic derivative used for constructing the global Hamiltonian becomes the usual derivative with a scaling factor, as is the case with standard BQISM.

Other boundary conditions have been studied for the Fateev--Zamolodchikov model. Boundaries were considered in \cite{Zhou1997} using the language of weights and the star-triangle relation. In the particular case of the original $D(D_{3})$ limit $K$-matrices were studied in \cite{DFIL2009}. \\



\noindent
\underline{\textbf{Braided closed boundary models}} \\
\noindent
To construct a braided model we set
$$ R_{12}'(\tilde{z}) = R_{21}(\tilde{z}), \hspace{0.5cm} \bar{R} = \lim_{\tilde{z} \rightarrow \tilde{z}_{0}} R'(\tilde{z}) \hspace{0.5cm} \mbox{and} \hspace{0.5cm} \bar{R}' = \lim_{\tilde{z} \rightarrow \tilde{z}_{0}} R(\tilde{z}), $$
where 
$$ \tilde{z}_{0} \in \{(0,0), (0,\infty), (\infty,0), (\infty,\infty)\}. $$
Using these we have Equation (\ref{YBE1}) implies Equations (\ref{YBE2}-\ref{YBE4}). Reflection matrices $K^{+}(\tilde{z})$ and $K^{-}(\tilde{z})$ are still required, however, given $K^{+}(\tilde{z})$ and $K^{-}(\tilde{z})$ which lead to an open model we can set
$$ K^{\pm} = \lim_{\tilde{z} \rightarrow \tilde{z}_{0}} K^{\pm}(\tilde{z}), $$
obtaining an integrable model. Here we will consider the case where $K^{\pm} = I$, this will lead to the global Hamiltonian
$$ \mathcal{H} = \sum_{i=1}^{\L -1}H_{i(i+1)} + GH_{(\L-1)\L}G^{-1}, $$
where $H$ is the local Hamiltonian described by Equation (\ref{Hlocal}),
$$ G = t(\tilde{1})  = b_{1}b_{2}...b_{\L-1} \, \mbox{tr}_{0}\left[P_{0\L}\bar{R}_{0\L}\right] \hspace{0.7cm} \mbox{and} \hspace{0.7cm} b_{i} = \bar{R}_{i(i+1)}P_{i(i+1)}. $$ 
For all choices of $z_{0}$ we have that $\mbox{tr}_{0}\left[P_{0\L}\bar{R}_{0\L}\right]$ is a scalar multiple of the identity and the global Hamiltonian is invariant under the action of $D(D_{n})$. Like the case with the twisted boundary model we have that $t(\tilde{1})$ again plays the role of a generalised translation operator. Here $t(\tilde{1})$ is a product of solutions to the braid equation,
$$ b_{i} b_{i+1} b_{i} = b_{i+1} b_{i} b_{i+1}. $$
Thus the model is invariant under translation through braiding.

We have four possibilities for $z_{0}$ but from unitarity we know that
$$ \lim_{\tilde{z} \rightarrow \tilde{z}_{0}} R'(\tilde{z}) P = \lim_{\tilde{z} \rightarrow \tilde{z}_{0}} \left[R'(\tilde{z}^{-1}) P\right]^{-1}, $$
thus the braiding operator $b_{i}$ generated from $(0,0)$ will be the inverse of the braiding operator from $(\infty,\infty)$. Likewise for the two remaining possibilities. If $\tilde{z}_{0} = (0,0)$ then we find that the braiding operator comes from the representation of the canonical element associated with $D(D_{n})$,
$$ b_{1} = P_{12} (\pi \tp \pi)\mathfrak{R}_{12} = \sum_{i,j=1}^{n} e_{i,i+j} \tp e_{i-j,i}. $$
In the case that $\tilde{z}_{0} = (0,\infty)$ the braiding operator is not of such an elegant form,
$$ b_{1} = \sum_{a,i,j,k=1}^{n} w^{4k(j-a-k) + 2j(a-j)} e_{i+a+j,i+a} \tp e_{i+j,i}. $$
For some other models it is possible to set $b_{1} = P$, in which case periodic chains appear.

We remark that in contrast to earlier work on braided models \cite{LinFoe1997} we have not assumed any properties of the $R$-matrix except for regularity. This modification was possible because the braided models can be viewed as modified open models and subsequently the original approaches retained many of Skylanin's unnecessary assumptions. If we further take this view then it appears regularity is also not necessary \cite{Zhou1996}.



\section{Descendant of the Six-Vertex model and Constructing Fusion Relations}
\noindent
The general $R$-matrix presented in \cite{FDIL2011} was constructed as a descendant of the zero-field six-vertex model while requiring the symmetry of $D(D_{n})$. Here we started with the Fateev--Zamolodchikov model, a known descendant of the zero-field six-vertex model, and took a special limit in which $D(D_{n})$ symmetry emerges. 
We use this connection to the six-vertex model to derive fusion relations.

The zero-field six-vertex $R$-matrix is given by
$$ r(z) = 
	\left(
	\begin{array}{cccc}
		w^{2}z^{-1} - w^{-2}z & 0 & 0 & 0\\
		0 & z^{-1} - z & w^{2} - w^{-2} & 0\\
		0 & w^{2} - w^{-2} & z^{-1} - z & 0\\
		0 & 0 & 0 & w^{2}z^{-1} - w^{-2}z
	\end{array}
\right), $$
and is associated with the $L$-operator
\begin{equation}
	L(z) = \sum_{k=0}^{n-1}\left \{ \left(w^{2k}e_{1,2} + w^{-2k}e_{2,1}\right) \tp e_{k,k} - iw^{-1}z \left[ e_{1,1} \tp e_{k-1,k} + e_{2,2} \tp e_{k+1,k} \right] \right\}. \nonumber 
\end{equation}
This operator is the same as that presented in \cite{FDIL2011} and is a special case of the $L$-operator presented in \cite{BazStr1990}, appearing in connection to the chiral Potts model. The former of these articles discusses the alteration of the underlying algebraic structure caused by taking the required limit.

These operators satisfy the Yang--Baxter like relations
\begin{eqnarray*}
	r_{12}(x) L_{13}(xy) L_{23}(y) & = & L_{23}(y) L_{13}(xy) r_{12}(x), \\
	\left[\lim_{z \rightarrow 0} z\, r_{12}(z) \right] L_{13}(x) L_{23}^{*}(y) & = & L_{23}^{*}(y) L_{13}(x) \left[\lim_{z\rightarrow 0} z\, r_{12}(z) \right], \\
	L_{12}(x_{1}) L_{13}(x_{1}y_{1}) R_{23}(y_{1},y_{2}) & = & R_{23}(y_{1},y_{2}) L_{13}(x_{1}y_{1}) L_{12}(x_{1}), \\
	L_{12}^{*}(x_{2}) L_{13}^{*}(x_{2}y_{2}) R_{23}(y_{1},y_{2}) & = & R_{23}(y_{1},y_{2}) L_{13}^{*}(x_{2}y_{2}) L_{12}^{*}(x_{2}).
\end{eqnarray*}
These relations can be used to construct commuting transfer matrices along with fusion relations.
In \cite{CDIL2010} fusion relations were presented for the original $D(D_{3})$ limit of the Fateev--Zamolodchikov model. Following their work we define the vectors
\begin{eqnarray*}
	v^{\pm}_{k} & = & \frac{1}{\sqrt{2}}\left[w^{k+\frac{n-1}{2}}v_{1}\tp v_{k} \pm w^{-k-\frac{n-1}{2}}v_{2}\tp v_{k-1}\right] \hspace{1cm} \mbox{and} \\
	u^{\pm}_{k} & = & \frac{1}{\sqrt{2}}\left[w^{k+\frac{n-1}{2}}v_{1}\tp v_{k-1} \pm w^{-k-\frac{n-1}{2}}v_{2}\tp v_{k}\right].
\end{eqnarray*}
Each set of vectors, $\{ v^{\pm}_{i} \}_{i=1}^{n}$ and $\{ u^{\pm}_{i} \}_{i=1}^{n}$, are orthonormal and satisfy
$$ L(z)v^{\pm}_{k} = -iw^{-1}[z \pm i] u^{\pm}_{k} \hspace{0.5cm} \mbox{which implies} \hspace{0.5cm} U^{-1}L(z)V = -iw^{-1}\left(\begin{array}{cc} [z+i] I_{n} & 0 \\ 0 & [z-i]I_{n} \end{array} \right), $$
for certain matrices $U$ and $V$ where $I_{n}$ is the identity $n$ by $n$ matrix.

Before we derive the fusion relations it useful to rescale the $R$-matrix, yielding
$$ R(\tilde{z}) = N(z_{1},z_{2})\sum_{a,i,j=1}^{n} \left[\sum_{b=1}^{n} w^{-2a(2b-j)} \overline{W}(z_{1}|b) \overline{W}(z_{2}^{-1}|b-j) \right] e_{i+j,i+a} \tp e_{i+a+j,i}, $$
where
$$ N(z_{1},z_{2}) = \frac{1}{n}\prod_{k=1}^{\frac{n-1}{2}}  (z_{1} - w^{4k}) (z_{2} - w^{-4k}). $$
This has the advantage that the entries in $R(\tilde{z})$ are now polynomial. This rescaling does not effect the global Hamiltonian due to the requirement that its trace is zero. \\

\noindent
\underline{\textbf{Periodic boundary models}} \\
\noindent
Looking at the Yang--Baxter like relations we find $L_{13}(iz_{1}) R_{23}(z_{1},z_{2})$ can be made lower block triangular. Algebraically we have found 
\begin{eqnarray*}
V_{12}^{-1}L_{13}(iz_1)R_{23}(z_1,z_2)V_{12}=\left(
\begin{array}{cc}
(w^{-2}z_1+1)R(w^{-2}z_1,z_2) & 0 \\
\star & (z_1 -1) R(w^{2}z_1,z_2)\end{array}\right),
\end{eqnarray*}
where $\star$ represents an unknown operator. This relation can be used to define a functional relation for models with periodic boundary conditions. 
\begin{eqnarray*}
	t^{(2)}(iz_{1})t^{(3)}(z_{1},z_{2})&=&(w^{-2}z_1+1)^{\L}t^{(3)}(w^{-2}z_1,z_2) + (z_1-1)^{\L}t^{(3)}(w^{2}z_1,z_2), \hspace{0.5cm} \mbox{where}\\
	t^{(2)}(z) & = & \mbox{tr}_{0} \left[ L_{0\L}(z)...L_{01}(z) \right], \\ 
	t^{(3)}(\tilde{z}) & = & \mbox{tr}_{0} \left[ R_{0\L}(\tilde{z})...R_{01}(\tilde{z}) \right].
\end{eqnarray*}
Here we have that the functional coefficients of the transfer matrices are only dependant upon $z_{1}$. It is possible to construct a second function relation where the functional coefficients are dependant upon $z_{2}$. In this case this is achieved by taking the complex conjugate of the above relation while noting that
$$ \left[t^{(3)}(z_{1},z_{2})\right]^{*} = t^{(3)}(z_{2}^{*},z_{1}^{*}). $$
For more general $t^{(3)}(\tilde{z})$, i.e. with twisted boundaries, this will not be case and the two functional relations which can be constructed will not be trivially related. Instead of using $\left[t^{(2)}(z)\right]^{*}$ another transfer matrix would need to be constructed using $L^{*}(z)$.

Using the commuting nature of the transfer matrices we write the functional relation as
$$ \lambda(iz_{1})\Lambda(z_{1},z_{2}) = (w^{-2}z_1+1)^{\L}\Lambda(w^{-2}z_1,z_2) + (z_1-1)^{\L}\Lambda(w^{2}z_1,z_2) $$
where $\lambda(z)$ and $\Lambda(\tilde{z})$ are eigenvalues of $t^{(2)}(z)$ and $t^{(3)}(\tilde{z})$ respectively. Following the work of $\cite{CDIL2010,FFL2011}$ we use the ansatz, $\Lambda(\tilde{z})$ is a constant or
\begin{equation} \label{EigAnsatz}
	\Lambda(\tilde{z}) = c \prod_{k=1}^{d_{1}}(z_{1}-iwy_{1,k})\prod_{k=1}^{d_{2}}(z_{2}+iw^{-1}y_{2,k}^{*}),
\end{equation}
where $d_{1}, d_{2} \in \N$ and $y_{1,k},y_{2,k} \in \C$. Dividing the functional relation through by $\Lambda(z_{1},z_{2})$ and taking the limit $z_{1} \rightarrow iwy_{1,j}$ it follows by looking at the residues that
$$ (-1)^{\L+1}\left(\frac{1+iw^{-1}y_{1,j}}{1-iwy_{1,j}}\right)^{\L} = \prod_{k=1}^{d_{1}} \left( \frac{y_{1,k}-w^{2}y_{1,j}}{y_{1,k}-w^{-2}y_{1,j}} \right), $$
for $1 \leq j \leq d_{1}$. It is likewise possible to construct a second Bethe equation from the second functional equation that was mentioned. The second equation in this case can be obtained by replacing $y_{1,k}$ with $y_{2,k}$ and $d_{1}$ with $d_{2}$ in the above equation. \\

\noindent
\underline{\textbf{Open and braided closed boundary models}} \\
\noindent
To construct fusion relations for models with open or braided closed boundary conditions we additionally require the operator
$$ L'(z) = \sum_{k=0}^{n-1} \left \{ \left[ e_{1,1} \tp e_{k+1,k} + e_{2,2} \tp e_{k-1,k} \right] - iw^{-1}z\left(w^{2k}e_{1,2} + w^{-2k}e_{2,1}\right) \tp e_{k,k} \right\} \propto L^{-1}(z^{-1}), $$
as $L(z)$ does not satisfy unitarity. It is again possible to diagonalise this operator using the property,
$$ L'(z) u^{\pm}_{k} = \mp i[z \pm i] v^{\pm}_{k} \hspace{0.5cm} \mbox{which implies} \hspace{0.5cm} V^{-1}L'(z)U = i\left(\begin{array}{cc} -[z+i] I_{n} & 0 \\ 0 & [z-i]I_{n} \end{array} \right). $$
Thus is possible to diagonalise both $L(z)$ and $L'(z)$ such that corresponding blocks have the same zeros. This allows the construction of fusion relations for both open boundary and braided closed boundary conditions. 

One example of functional relations for the open model is
\begin{eqnarray*}
	t^{(2)}(iz_{1})t^{(3)}(z_{1},z_{2})&=& f(w^{-1}z_{1})t^{(3)}(w^{-2}z_1,z_2) + g(w^{-1}z_{1}) t^{(3)}(w^{2}z_1,z_2), \hspace{0.5cm} \mbox{where}\\
	t^{(2)}(z) & = & \mbox{tr}_{0} \left[ L_{0\L}(z)...L_{01}(z) L_{01}'(z)...L_{0\L}'(z) \right], \\ 
	t^{(3)}(\tilde{z}) & = & \mbox{tr}_{0} \left[ R_{0\L}(\tilde{z})...R_{01}(\tilde{z}) R_{10}(\tilde{z})...R_{\L 1}(\tilde{z}) \right] \\
	f(z) & = & \frac{(1+w^{2}z^{2})(1-w^{-2}z^{2})}{(1-z^{4})} [w^{-1}z+1]^{2\L},\\
	g(z) & = & (-1)^{\L}\frac{(1-w^{2}z^{2})(1+w^{-2}z^{2})}{(1-z^{4})}[wz-1]^{2\L},
\end{eqnarray*}
Here $t^{(3)}(\tilde{z})$ will produce an open quantum chain with non-interacting boundary terms. Again taking the conjugate equation of this equation yields and additional functional relation whose functional coefficients depend upon $z_{2}$. Incorporating interacting boundary terms will make the two functional relations non-trivially related. Using the ansatz given in Equation (\ref{EigAnsatz}) for the eigenvalues of $t^{(3)}(z_{1},z_{2})$ yields the Bethe equations
$$ (-1)^{\L+1} \left( \frac{1-w^{2}y_{a,j}^{2}}{1-w^{-2}y_{a,j}^{2}} \right) \left( \frac{1+w^{-2}y_{a,j}^{2}}{1+w^{2}y_{a,j}^{2}} \right) \left(\frac{1+iw^{-1}y_{a,j}}{1-iwy_{a,j}}\right)^{2\L} = \prod_{k=1}^{d_{a}}\left(\frac{y_{a,k}-w^{2}y_{a,j}}{y_{a,k}-w^{-2}y_{a,j}} \right), $$
for $a\in \{1,2\}$ and $1 \leq j \leq d_{a}$.

We also consider the braided closed boundary models. It is possible to consider the four different possibilities outline previously. We let
$$ \tilde{z}_{0} = (z_{01}, z_{02}) \in \{(0,0), (0,\infty), (\infty,0), (\infty,\infty)\} $$
and define the operators
$$ \bar{L}(z) = \lim_{x\rightarrow z} \left( \frac{1}{1+x} L'(x) \right) \hspace{0.5cm} \mbox{and} \hspace{0.5cm} \bar{R} = \lim_{z\rightarrow \tilde{z}_{0}} \left[ \left(1+z_{1}^{\frac{n-1}{2}}\right)^{-1} \left(1+z_{1}^{\frac{n-1}{2}}\right)^{-1} R_{21}(\tilde{z}) \right]. $$
We also make use of the following constants
$$ b_{0} = 1, \hspace{0.5cm} b_{\infty}=-iw^{-1}, \hspace{0.5cm} c_{0}=1 \hspace{0.5cm} \mbox{and} \hspace{0.5cm} c_{\infty}=iw^{-1}. $$
With these constants we have the following functional relations,
\begin{eqnarray*}
	t^{(2,1)}(iz_{1})t^{(3)}(z_{1},z_{2})&=& b_{z_{01}}^{\L} (w^{-2}z_{1}+1)^{\L} t^{(3)}(w^{-2}z_1,z_2) + c_{z_{01}}^{\L} (z_{1}-1)^{\L} t^{(3)}(w^{2}z_1,z_2), \\
	t^{(2,2)}(-iz_{2})t^{(3)}(z_{1},z_{2})&=& (b_{z_{02}}^{*})^{\L} (w^{2}z_{2}+1)^{\L} t^{(3)}(z_1,w^{2}z_2) + (c_{z_{02}}^{*})^{\L} (z_{2}-1)^{\L} t^{(3)}(z_1,w^{-2}z_2), \\
	t^{(2,1)}(z) & = & \mbox{tr}_{0} \left[ L_{0\L}(z)...L_{01}(z) \bar{L}_{01}(z_{01})...\bar{L}_{0\L}(z_{01}) \right], \\
	t^{(2,2)}(z) & = & \mbox{tr}_{0} \left[ L_{0\L}^{*}(z)...L_{01}^{*}(z) \bar{L}_{01}^{*}(z_{02})...\bar{L}_{0\L}^{*}(z_{02}) \right], \\ 
	t^{(3)}(\tilde{z}) & = & \mbox{tr}_{0} \left[ R_{0\L}(\tilde{z})...R_{01}(\tilde{z}) \bar{R}_{01}...\bar{R}_{1\L} \right].
\end{eqnarray*}
Here we note that the two functional equations are non-trivially related. Using these functional relations and the ansatz presented in equation (\ref{EigAnsatz}) we have the Bethe equation
$$ (-1)^{\L+1}\left(\frac{b_{z_{0a}}}{c_{z_{0a}}} \right)^{\L} \left(\frac{1+iw^{-1}y_{a,j}}{1-iwy_{a,j}}\right)^{\L} = \prod_{k=1}^{d_{a}} \left( \frac{y_{a,k}-w^{2}y_{a,j}}{y_{a,k}-w^{-2}y_{a,j}} \right), $$
for $a\in \{1,2\}$ and $1 \leq j \leq d_{a}$. These equations at most differ by a factor of $(-1)^{\L}$ on the left hand side of the equation compared to the periodic boundary case. \\

\section{Summary}
Taking a special case of the Fateev--Zamolodchikov model yields an $R$-matrix which intertwines the coproduct of $D(D_{n})$. Subsequently models with periodic, open or braided closed boundary conditions will have the underlying symmetry of $D(D_{n})$ (or one of its subalgebras). The Hamiltonians presented are seen to be composed of two Hamiltonians with a free coupling parameter. Functional equations were also constructed through fusion which allowed for Bethe equations to be determined. In the simplest case of $n=3$ the ground state energies of the periodic and open boundary models have been explicitly computed \cite{FFL2011} along with one case of the braided closed boundary models. \\

\noindent
\underline{\textbf{Acknowledgements}} \\
The author would like to thank Jon Links for much discussion and advice.


\noindent





\begin{thebibliography}{10}

\bibitem{Albertini1992}
G. Albertini, \textit{Bethe-ansatz type equations for the Fateev--Zamolodchikov
  spin model}, J. Phys. A: Math. Gen., \textbf{25}, 1799-1813, (1992).

\bibitem{AMCPT1989}
G. Albertini, B.M. McCoy, J.H.H. Perk and S. Tang, \textit{Excitation spectrum
  and order parameter for the integrable $N$-state chiral Potts model}, Nucl.
  Phys. B, \textbf{314}, 741--763, (1989).

\bibitem{AYMCPTY1987}
H. Au-Yang, B.M. McCoy, J.H.H. Perk, S. Tang and M. Yan, \textit{Commuting
  transfer matrices in the chiral Potts models: Solutions of star-triangle
  equations with genus$>$1}, Phys. Lett. A, \textbf{123}, 219--223, (1987).

\bibitem{AuYPer1997}
H. Au-Yang and J.H.H. Perk, \textit{The many faces of the chiral Potts model},
  Int. Journal of Mod. Phys. B, \textbf{11}, 11-26, (1997).

\bibitem{BaxterBook1982}
R.J. Baxter, \textit{Exactly solved models in statistical mechanics}, Academic
  Press, London, (1982).

\bibitem{BPAY1988}
R.J. Baxter, J.H.H. Perk and H. Au-Yang, \textit{New solutions of the
  star-triangle relations for the chiral Potts model}, Phys. Lett. A,
  \textbf{128}, 138--142, (1988).

\bibitem{BazPer2009}
V.V. Bazhanov and J.H.H. Perk, \textit{Connection of $D(D_{3})$ model and
  3-state self-dual Potts model I-II}, Private Communications, (2009).

\bibitem{BazStr1990}
V.V. Bazhanov and Y.G. Stroganov, \textit{Chiral Potts model as a descendant of
  the six-vertex model}, J. Stat. Phys., \textbf{59}, 799--817, (1990).

\bibitem{BGZZ1998}
A.J. Bracken, X.Y. Ge, Y.Z. Zhang and H.Q. Zhou, \textit{An open-boundary
  integrable model of three coupled {$XY$} spin chains}, Nucl. Phys. B,
  \textbf{516}, 603--622, (1998).

\bibitem{CDIL2010}
C.W. Campbell, K.A. Dancer, P.S. Isaac and J. Links, \textit{Bethe ansatz
  solution of an integrable, non-Abelian anyon chain with $D(D_3)$ symmetry},
  Nucl. Phys. B, \textbf{836}, 171-185, (2010).

\bibitem{CGX1991}
Y. Cheng, M.L. Ge and K. Xue, \textit{Yang--Baxterisation of braid group
  representations}, Commun. Math. Phys., \textbf{136}, 195--208, (1991).

\bibitem{DFIL2009}
K.A. Dancer, P.E. Finch, P.S. Isaac and J. Links, \textit{Integrable boundary
  conditions for a non-Abelian anyon chain with $D(D_{3})$ symmetry}, Nucl.
  Phys. B, \textbf{812}, 456--469, (2009).

\bibitem{DIL2006}
K.A. Dancer, P.S. Isaac and J. Links, \textit{Representations of the quantum
  double of finite group algebras and spectral parameter dependent solutions of
  the Yang--Baxter equation}, J. Math. Phys., \textbf{47}, 103511, (2006).

\bibitem{Vega1984}
H.J. {de Vega}, \textit{Families of commuting transfer matrices and integrable
  models with disorder}, Nucl. Phys. B, \textbf{26}, 495-513, (1984).

\bibitem{DPR1990}
R. Dijkgraaf, V. Pasquier and P. Roche, \textit{Quasi Hopf algebras, group
  cohomology and orbifold models}, Nucl. Phys. (Proc. Supp.), \textbf{18},
  60--72, (1990).

\bibitem{FatZam1982b}
V.A. Fateev and A.B. Zamolodchikov, \textit{Self-dual solutions of the
  star-triangle relation in $\mathbb{Z}_{N}$-Models}, Phys. Lett. A,
  \textbf{92}, 37--39, (1982).

\bibitem{FTLTKWF2007}
A. Feiguin, S. Trebst, A.W.W. Ludwig, M. Troyer, A.Y. Kitaev, Z. Wang and M.H.
  Freedman, \textit{Interacting Anyons in Topological Quantum Liquids: The
  Golden Chain}, Physical Review Letters, \textbf{98}, 160409, (2007).

\bibitem{FDIL2011}
P.E. Finch, K.A. Dancer, P.S. Isaac and J. Links, \textit{Solutions of the
  Yang-Baxter equation: descendants of the six-vertex model from the Drinfeld
  doubles of dihedral group algebras}, Nucl. Phys. B, \textbf{847}, 387--412,
  (2011).

\bibitem{FFL2011}
P.E. Finch, H. Frahm and J. Links, \textit{Ground-state phase diagram for a
  system of interacting, $D(D_3)$ non-Abelian anyons}, Nucl. Phys. B,
  \textbf{844}, 129--145, (2011).

\bibitem{Foerster1996}
A. Foerster, \textit{Quantum group invariant supersymmetric {$t$}-{$J$} model
  with periodic boundary conditions}, J. Phys. A, \textbf{29}, 7625--7633,
  (1996).

\bibitem{FLR1998}
A. Foerster, J. Links and I. Roditi, \textit{Integrable multiparametric quantum
  spin chains}, J. Phys. A: Math. Gen, \textbf{31}, 687-695, (1998).

\bibitem{FraleighBook1982}
J.B. Fraleigh, \textit{A first course in abstract algebra (3rd ed.)},
  Addison-Wesley, (1982).

\bibitem{Gould1993}
M.D. Gould, \textit{Quantum double finite group algebras and their
  representations}, Bull. Aust. Math. Soc., \textbf{48}, 275--301, (1993).

\bibitem{GPPR1994}
H. Grosse, S. Pallua, P. Prester and E. Raschhofer, \textit{On a quantum group
  invariant spin chain with non-local boundary conditions}, J. Phys. A: Math.
  Gen., \textbf{27}, 4761-4771, (1994).

\bibitem{Jones1990}
V.F.R. Jones, \textit{Baxterization}, Int. J. Mod. Phys. B, \textbf{4},
  701--713, (1990).

\bibitem{KarZap1994}
M. Karowski and A. Zapletal, \textit{Quantum Group Invariant Integrable n-State
  Vertex Models with Periodic Boundary Conditions}, Nucl. Phys. B,
  \textbf{419}, 567-588, (1994).

\bibitem{LimaSantos1998}
A. Lima-Santos, \textit{Exact solutions of graded {T}emperley-{L}ieb
  {H}amiltonians}, Nuclear Phys. B, \textbf{522}, 503--532, (1998).

\bibitem{LinFoe1997}
J. Links and A. Foerster, \textit{On the construction of integrable closed
  chains with quantum supersymmetry}, J. Phys. A, \textbf{30}, 2483--2487,
  (1997).

\bibitem{LFK1999}
J. Links, A. Foerster and M. Karowski, \textit{Bethe ansatz solution of a
  closed spin {$1\ XXZ$} {H}eisenberg chain with quantum algebra symmetry}, J.
  Math. Phys., \textbf{40}, 726--735, (1999).

\bibitem{LinGou1996}
J. Links and M.D. Gould, \textit{Integrable systems on open chains with quantum
  supersymmetry}, Int. J. Mod. Phys. B, \textbf{25}, 3461-3480, (1996).

\bibitem{McGuire1966}
J.B. McGuire, \textit{Interacting fermions in one dimension. II. Attractive
  potential}, J. Math. Phys., \textbf{7}, 123, (1966).

\bibitem{MezNep1991}
L. Mezincescu and R.I. Nepomechie, \textit{Integrable open spin chains with
  non-symmetric $R$-matrices}, J. Phys. A: Math. Gen., \textbf{24}, L15-L23,
  (1991).

\bibitem{MezNep1992}
L. Mezincescu and R.I. Nepomechie, \textit{Addendum: Integrability of open
  chains with quantum algebra symmetry}, Int. J. Mod. Phys. A, \textbf{7},
  5657-5659, (1992).

\bibitem{NSSFDS2008}
C. Nayak, S.H. Simon, A. Stern, M. Freedman and S. {Das Sarma},
  \textit{Non-Abelian Anyons and topological quantum computation}, Rev. Mod.
  Phys., \textbf{80}, 1083, (2008).

\bibitem{PerAuY2006}
J.H.H. Perk and H. Au-Yang, \textit{The Yang--Baxter equations}, Encyclopedia
  of Mathematical Physics, Eds. J.-P. Françoise, G.L. Naber and S.T. Tsou
  (Oxford: Elsevier), \textbf{5}, 465, (2006).

\bibitem{RaySha2005}
S. Ray and J. Shamanna, \textit{A Bethe ansatz study of free energy and
  excitation spectrum for even spin Fateev--Zamolodchikov model}, J. Math.
  Phys., \textbf{46}, 043301, (2005).

\bibitem{ShaSut1990}
B.S. Shastry and B. Sutherland, \textit{Twisted boundary conditions and
  effective mass in Heisenberg-Ising and Hubbard rings}, Physical review
  letters, \textbf{65}, 243–246, (1990).

\bibitem{TTWL2009}
M.T. Simon Trebst~and, Z. Wang and A.W.W. Ludwig, \textit{A short introduction
  to Fibonacci anyon models}, arXiv:0902.3275v1, (2009).

\bibitem{Sklyanin1988}
E.K. Sklyanin, \textit{Boundary conditions for integrable quantum systems}, J.
  Phys. A: Math. Gen., \textbf{21}, 2375-2389, (1988).

\bibitem{STF1979}
E.K. Sklyanin, L.A. Takhtadzhyan and L.D. Faddeev, \textit{Quantum inverse
  problem method. 1}, Theor. Math. Phys., \textbf{40}, 688-706, (1979).

\bibitem{Yang1967}
C.N. Yang, \textit{Some exact results for the many-body problem in one
  dimension with repulsive delta-function interaction}, Phys. Rev. Lett.,
  \textbf{19}, 1312--1315, (1967).

\bibitem{Zhou1996}
H.Q. Zhou, \textit{Integrable open-boundary conditions for the one-dimensional
  Bariev chain}, Phys. Rev. B, \textbf{53}, 5098--5100, (1996).

\bibitem{Zhou1997}
Y.K. Zhou, \textit{Fateev--Zamolodchikov and Kashiwara-Miwa models: boundary
  star-triangle relations and surface critical properties}, Nucl. Phys. B,
  \textbf{487}, 779--794, (1997).

\end{thebibliography}
\end{document}